\documentclass{article}[12pt]
\usepackage{amsmath,amssymb}
\usepackage{graphicx}%
\usepackage{verbatim}

\newcommand{\TwoFig}[4]{%
\begin{center}
\begin{tabular}{lr}
\parbox{8cm}{\includegraphics[width=8cm]{#1}}  & \parbox{8cm}{\includegraphics[width=8cm]{#2}} \\
\parbox{8cm}{\vspace{7pt}\refstepcounter{figure}Figure \thefigure.\quad #3\vfill} & \parbox{8cm}{\vspace{7pt}\refstepcounter{figure}Figure \thefigure.\quad #4\vfill} \\
\end{tabular}
\end{center}
\vspace{12pt}
}

\begin{document}

\begin{center}
{\bf \Large COMPLETE COSMOLOGICAL EVOLUTION MODEL OF A CLASSICAL SCALAR FIELD WITH A HIGGS POTENTIAL. II. NUMERICAL SIMULATION\\[12pt]
Yu. G. Ignat'ev, A. R. Samigullina }\\
Physics Institute of Kazan Federal University,\\
Kremleovskaya str., 35, Kazan, 420008, Russia.
\end{center}

\begin{abstract}

A complete cosmological evolution model of the classical scalar field with a Higgs potential is studied and simulated on a computer without the assumption that the Hubble constant is nonnegative. It is shown that with most initial conditions, the cosmological model goes from the expansion stage to the compression stage. Thus, cosmological models based on the classical Higgs field are unstable to finite perturbations. \\

{\bf keywords}: cosmological models, Higgs fields, Einstein-Higgs hypersurface, global behavior.

\end{abstract}
\section{DYNAMIC SYSTEM AND COMMENTS ON THE CAUCHY PROBLEM}
In the first part of our work \cite{IgnSamig}, the basic mathematical relationships
were formulated for the cosmological model based on the classical scalar Higgs field.
First, they include the normal independent system of differential equations
\begin{equation} \label{Eq__1}
\Phi'=Z,
\end{equation}
\begin{equation} \label{Eq__2}
Z'=-3hZ-e\Phi+\alpha_m\Phi^3,
\end{equation}
\begin{equation} \label{Eq__3}
Z'=-3h^2+\frac{e\Phi^2}{2}-\frac{\alpha_m\Phi^4}{4}+\lambda_m,
\end{equation}
where the derivatives with respect to time $\tau=mt$  are in Compton units $\lambda=\lambda_0-m^4/4\alpha_m$ , and other designations can be found in \cite{IgnSamig}. Second, they involve the Einstein-Higgs hypersurface equation, being, as a matter of fact, on the one hand, the Einstein equation $^4_4$, and on the other hand, the first integral of dynamic system \eqref{Eq__1}-\eqref{Eq__3}:
\begin{equation} \label{Eq__4}
-3h^2-\frac{Z^2}{2}-\frac{e\Phi^2}{2}+\frac{\alpha_m\Phi^4}{4}-\lambda_m=0,
\end{equation}
Third, they involve relationships for the physical characteristics of the cosmological models, including
the invariant cosmological acceleration 	
\begin{equation} \label{Eq__5}
\Omega\equiv\frac{a\ddot{a}}{\dot{a}^2}\equiv 1+\frac{\dot{H}}{H^2}
\end{equation}
the invariant curvature of the Friedman space
\begin{equation} \label{Eq__6}
\sigma\equiv\sqrt{R_{ijkl}R^{ijkl}}=H^2\sqrt{6(1+\Omega^2)}\equiv\sqrt{6}\sqrt{H^4+(H^2+\dot{H})^2}\geqslant 0,
\end{equation}
and the effective energy of the dynamic system
\begin{equation} \label{Eq__7}
\mathcal{E}_m(\Phi,Z)=\frac{Z^2}{2}+\frac{e\Phi^2}{2}-\frac{\alpha_m\Phi^4}{4}+\lambda_m\geqslant 0
\end{equation}
And fourth, they comprise some differential relationships between the dynamic variables that are a consequence of the complete system of the Einstein equations and the scalar Higgs field, in particular, the equivalent form of \eqref{Eq__5}
\begin{equation} \label{Eq__8}
h'=-\frac{Z^2}{2}(\leqslant 0),
\end{equation}
according to which the Hubble constant cannot increase with time in cosmological models with the classical scalar Higgs field.
Turning to numerical simulation of the dynamic system under study, we note first that some special
features in the behavior of this system have already been revealed by numerical simulation in \cite{Ign}.
Following this work, we define dynamic system \eqref{Eq__1}-\eqref{Eq__3} by the ordered list of three dimensionless
 parameters $\textbf{P}=[e,\alpha_m,\lambda_m]$  and the ordered list of initial conditions $\textbf{I}=[\Phi_0,Z_0,\varepsilon]$ ,
where $\varepsilon=\pm 1$, $\varepsilon=+1$,   corresponds the expansion phase at the initial time $t_0$, and $\varepsilon=-1$
corresponds to the compression phase at this time moment. According to the foregoing, the initial value of the Hubble constant is defined by Einstein equation \eqref{Eq__4} from which we obtain
\begin{equation} \label{Eq__9}
h_0=\pm \frac{1}{3}\sqrt{\frac{Z^2_0}{2}+\frac{e\Phi^2_0}{2}-\frac{\alpha_m\Phi^4_0}{4}+\lambda_m}\equiv\frac{\varepsilon}{3} \sqrt{\mathcal{E}^0_m},
\end{equation}
Second, we note that dynamic system \eqref{Eq__1}-\eqref{Eq__3} is the system of ordinary differential equations explicitly independent of the time variable and hence invariant to the translation $t\rightarrow t_0+t$  \cite{Ign}. Therefore, any arbitrary  $t_0$ value can be chosen as the initial time moment for the initial conditions. We set this value equal to zero. Moreover, we can consider the states of the dynamic system for negative times $t_0<0$ .
\section{RESULTS OF NUMERICAL SIMULATION}
\subsection{Dynamic system parameters $P=[1, 1, 0.3]$}
Figure 1 shows the card of singular points of the dynamic system. Hereinafter, closed circles show attractive foci (af), asterisks show repulsive foci (rf), and diagonal crosses show saddle points (sd). Figure 2 shows the Einstein-Higgs surface for the indicated set of the parameters.

Thus, in this case there are all six special points: $[[0,-0.3162277660], af]$, $[[0, 0.3162277660], rf]$,\\ $[[-1, -0.4281744192], sd]$, $[[1,-0.4281744192], sd], [[-1, 0.4281744192], sd]$, and $[[1, 0.4281744192], sd]$, two of which are foci - attractive (at the bottom of the figure) and repulsive (at the top of the figure) - and four are saddle points (at the corners of the figure).
Fig. 1. Card of singular points of dynamic system \eqref{Eq__1}-\eqref{Eq__3} with the parameters $\textbf{P} = [1, 1, 0.3]$.
Fig. 2. Einstein-Higgs hypersurface of dynamic system \eqref{Eq__1}-\eqref{Eq__3} with the parameters $\textbf{P} = [1, 1, 0.3]$.
According to the classification of the Einstein-Higgs hypersurfaces (see \cite{IgnSamig}), the Einstein-Higgs hypersurface in this case represents a deformed single-sheet hyperboloid with the OZ major axis. Thus, the topology of the Einstein hypersurface admits rolling of the phase trajectories from the top down the surface. Figures 3 and 4 illustrate the dependences   for different initial conditions (Fig. 3) and the phase trajectories of the dynamic system in the   plane against the background of the card of singular points and in the cross section of the Einstein-Higgs hypersurface   (Fig. 4).
\TwoFig{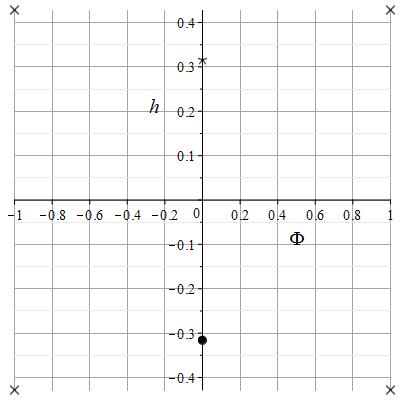}{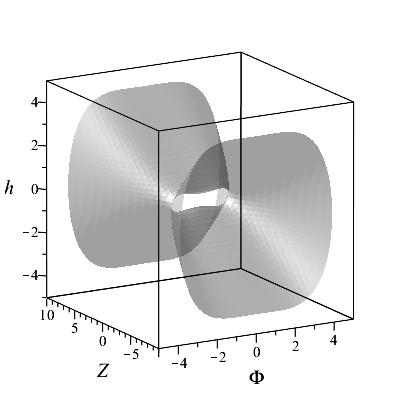}{Card of singular points of dynamic system (1)-(3) with the parameters $\textbf{P}=[1, 1, 0.3]$ \label{Fig_1}}{Einstein-Higgs hypersurface of dynamic system (1)-(3) with the parameters $\textbf{P}=[1, 1, 0.3]$}
\TwoFig{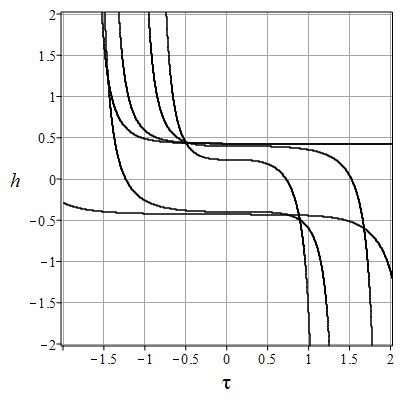}{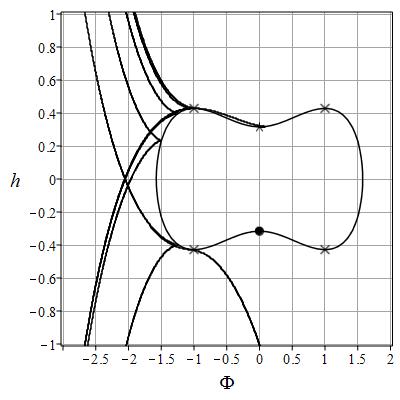}{Evolution of the Hubble constant of dynamic system \eqref{Eq__1}-\eqref{Eq__3} with the parameters $\textbf{P}=[1,1,0.3]$ \label{Fig_3}}
{Phase trajectories of dynamic system \eqref{Eq__1}-\eqref{Eq__3} with the parameters $\textbf{P}=[1,1,0.3]$ in the plane $S:Z=0$.}

Finally, Fig. 5 shows the phase trajectories of the dynamic system superimposed on its Einstein-Higgs hypersurface, and Fig. 6 shows the evolution of the invariant acceleration.
\TwoFig{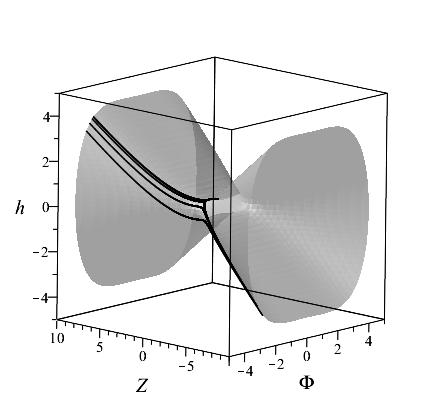}{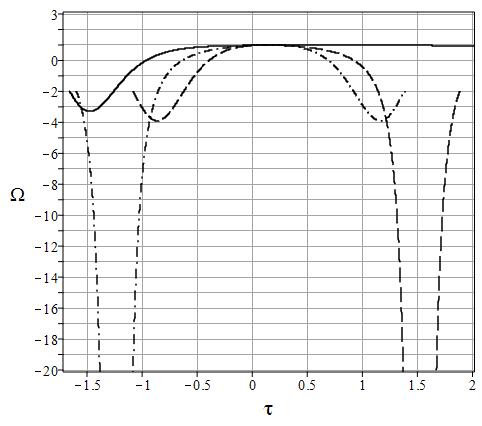}{Phase trajectories of dynamic system \eqref{Eq__1}-\eqref{Eq__3}
with the parameters $\textbf{P}=[1, 1, 0.3]$ on the Einstein-Higgs hypersurface.}{Invariant acceleration of dynamic system
\eqref{Eq__1}-\eqref{Eq__3} with the parameters $\textbf{P}=[1, 1, 0.3]$
and the initial conditions $\textbf{I}=[-1, 0.1, 1]$ (the solid curve),
$[-1.25, 0.1, 1]$ (the dashed curve), and $[-1.25, 0.1,-1]$ (the dot-dashed curve).}
\subsection{Dynamic system parameters $\textbf{P}=[-1,-1, 0.1]$}
Figure 7 shows the card of singular points of the dynamic system. Figure 8 shows the Einstein-Higgs surface for this set of the parameters.
\TwoFig{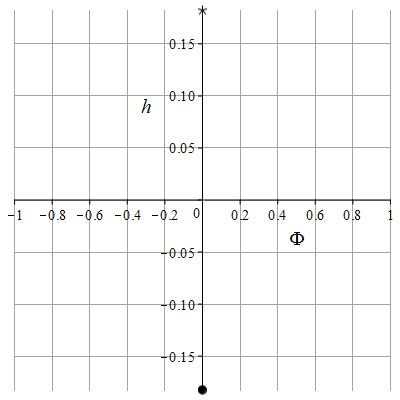}{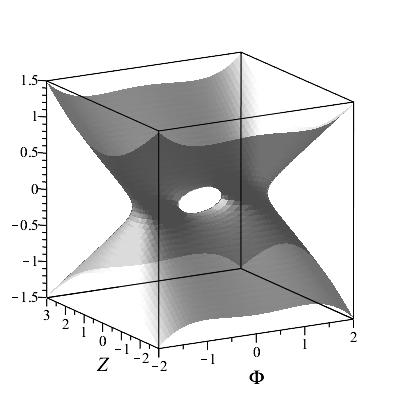}{Card of singular points of dynamic system \eqref{Eq__1}-\eqref{Eq__3} with the parameters
 $\textbf{P}=[-1,-1, 0.1]$.}{Einstein-Higgs surface of dynamic system \eqref{Eq__1}-\eqref{Eq__3}
with the parameters $\textbf{P}=[-1,-1, 0.1]$.}
In this case, there are only two singular points: $[[0,-0.1825741858], rf]$ and $[[0, 0.1825741858]$, af], one of which - the attraction center - is located at the bottom of the figure, and another - the repulsion center - is located at the top of the figure. The Einstein-Higgs hypersurface represents a deformed single-sheet hyperboloid with the \texttt{Oh} major axis. Thus, the topology of the Einstein hypersurface admits rolling of the phase trajectories from the top down the surface.

Figures 9 and 10 show the dependences $h(\tau)$  with different initial conditions (Fig. 9) and the phase trajectories of the dynamic system in the plane $\{\Phi,h\}$  against the background of the card of singular points and the cross section of the Einstein-Higgs hypersurface   (Fig. 10). Finally, Fig. 11 shows the phase trajectories of the dynamic system superimposed on its Einstein-Higgs hypersurface, and  Fig. 12 shows the evolution of the invariant acceleration.
\TwoFig{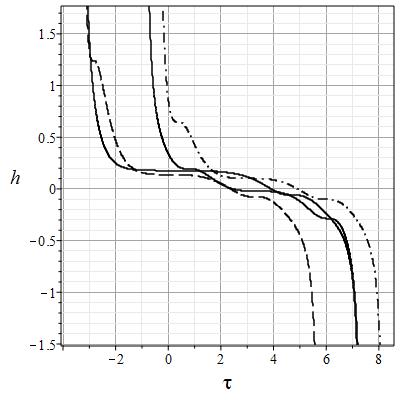}{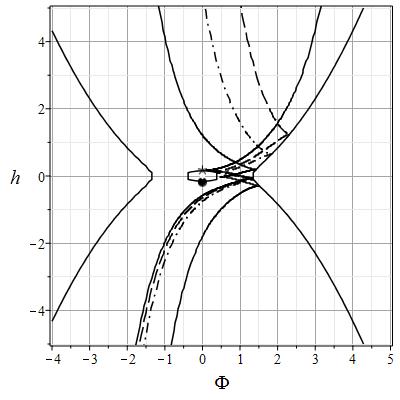}{Evolution of the Hubble constant of dynamic system \eqref{Eq__1}-\eqref{Eq__3} with the parameters $\textbf{P}=[-1,-1, 0.1]$.}{Phase trajectories of dynamic system \eqref{Eq__1}-\eqref{Eq__3} with the parameters $\textbf{P}=[-1,-1, 0.1]$ in the plane   against the background of the card of singular points.}
\TwoFig{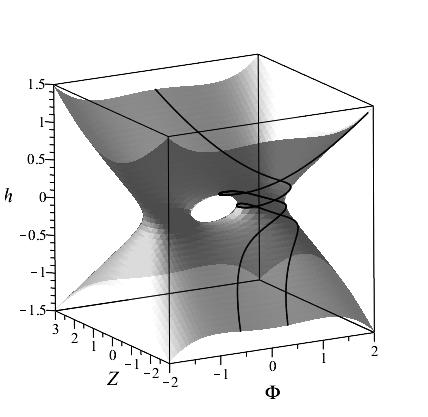}{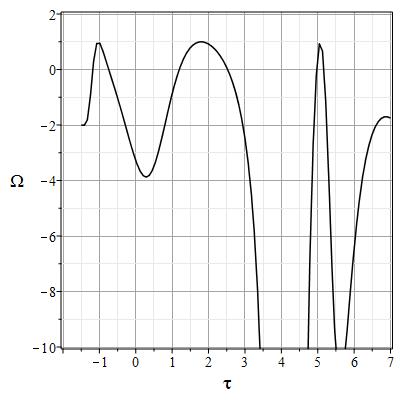}{ Phase trajectories of dynamic system \eqref{Eq__1}-\eqref{Eq__3} with the parameters $\textbf{P}=[-1,-1, 0.1]$ on the Einstein-Higgs hypersurface.}{Invariant acceleration of dynamic systems \eqref{Eq__1}-\eqref{Eq__3} with the parameters $\textbf{P}=[-1,-1, 0.1]$ for the initial conditions $\textbf{I}=[-1, 0.1, 1]$.}
\section{CONCLUSIONS}
In conclusion, we list the main results of the article. 
\begin{itemize}
  \item The cosmological models with the classical Higgs field show the tendency to go from the expansion mode  ($H>0$) to the compression mode  ($H<0$ )for a wide range of fundamental parameters and initial conditions;
  \item There are two rolling down modes $H_+\rightarrow H_-$ depending on the geometry of the Einstein hypersurface, more precisely, on the direction of its major axis.
  \item Constant solutions $\Phi=const$ normally used in standard cosmology correspond to the stable singular points (the attracting foci). However, for small finite deviation from the singular point, the solution rolls down infinite compression.
\item Thus, it is possible to state that the solutions used in standard scenarios yield small phase flux in comparison with the total one, that is, the probability of such solutions is low.
\end{itemize}

This paper was supported by the Kazan Federal University Strategic Academic Leadership Program.

\end{document}